\documentclass[aps,pra,twocolumn,amsmath,amssymb,superscriptaddress]{revtex4-1}
\usepackage{braket,graphicx, subfigure}

\begin{document}

\author{Alan O. Jamison}
\email{jamisona@uw.edu}
\affiliation{University of Washington Department of Physics, Seattle, Washington 98195, USA}

\author{Benjamin Plotkin-Swing}
\affiliation{University of Washington Department of Physics, Seattle, Washington 98195, USA}

\author{Subhadeep Gupta}
\affiliation{University of Washington Department of Physics, Seattle, Washington 98195, USA}

\date{\today}

\begin{abstract}
Using a three-path contrast interferometer (CI) geometry and laser-pulse diffraction gratings, we create the first matter-wave interferometer with ytterbium (Yb) atoms. We present advances in contrast interferometry relevant to high-precision measurements. By comparing to a traditional atom interferometer, we demonstrate the immunity of the CI to vibrations for long interaction times ($> 20\;{\rm ms}$). We characterize and demonstrate control over the two largest systematic effects for a high-precision measurement of the fine structure constant via photon recoil with our interferometer: diffraction phases and atomic interactions. Diffraction phases are an important systematic for most interferometers using large-momentum transfer beam splitters; atomic interactions are a key concern for any BEC interferometer. Finally, we consider the prospects for a future sub-part per billion photon recoil measurement using a Yb CI.
\end{abstract}

\title{Advances in precision contrast interferometry with Yb Bose-Einstein condensates}
\maketitle
\section{Introduction}
Many of the most surprising results of quantum theory arise from interference effects in the wave aspects of material particles. These same effects can be harnessed in a matter-wave interferometer for precision measurements \cite{cronin09}. Matter-wave interferometers have been used for a variety of precision measurements, from applications such as measuring gravity and gravity gradients\cite{sorrentino12} or rotation sensing\cite{durfee06} to fundamental physics such as measuring the fine structure constant\cite{bouchendira11} or atomic polarizabilities\cite{holmgren10}. Most precision measurements rely on incoherent sources of atoms such as beam lines or laser-cooled clouds. However, Bose-Einstein condensates (BECs) have recently received attention for a variety of interferometric measurements because their coherence properties have the potential to greatly enhance signal strength and visibility\cite{chiow11,hardman13}.

Atom interferometry has historically focused on the alkali atoms. In this paper, we report the first matter-wave interferometer using ytterbium (Yb). Unlike alkalis, the bosonic isotopes of Yb have no magnetic moment in the ground state. Eliminating the need for magnetic shielding makes Yb a promising atom for precision measurements\cite{dickerson12}. The large number of stable isotopes, both fermionic and bosonic\cite{fukuhara07}, allow a variety of properties to be modified between experiments, further enhancing the appeal of Yb for precision interferometry. Unlike alkali atoms, Yb has several transitions in the visible spectrum, including a strong dipole transition, a weak intercombination transition, and two clock transitions. This variety of transitions is key for proposed applications of Yb in interferometry\cite{graham13}\footnote{The current work may be extended to use different transitions for diffraction (intercombination transition) and readout light (strong dipole transition) with the potential to substantially increase signal to noise.}.

\begin{figure}
\includegraphics[width=0.9\linewidth]{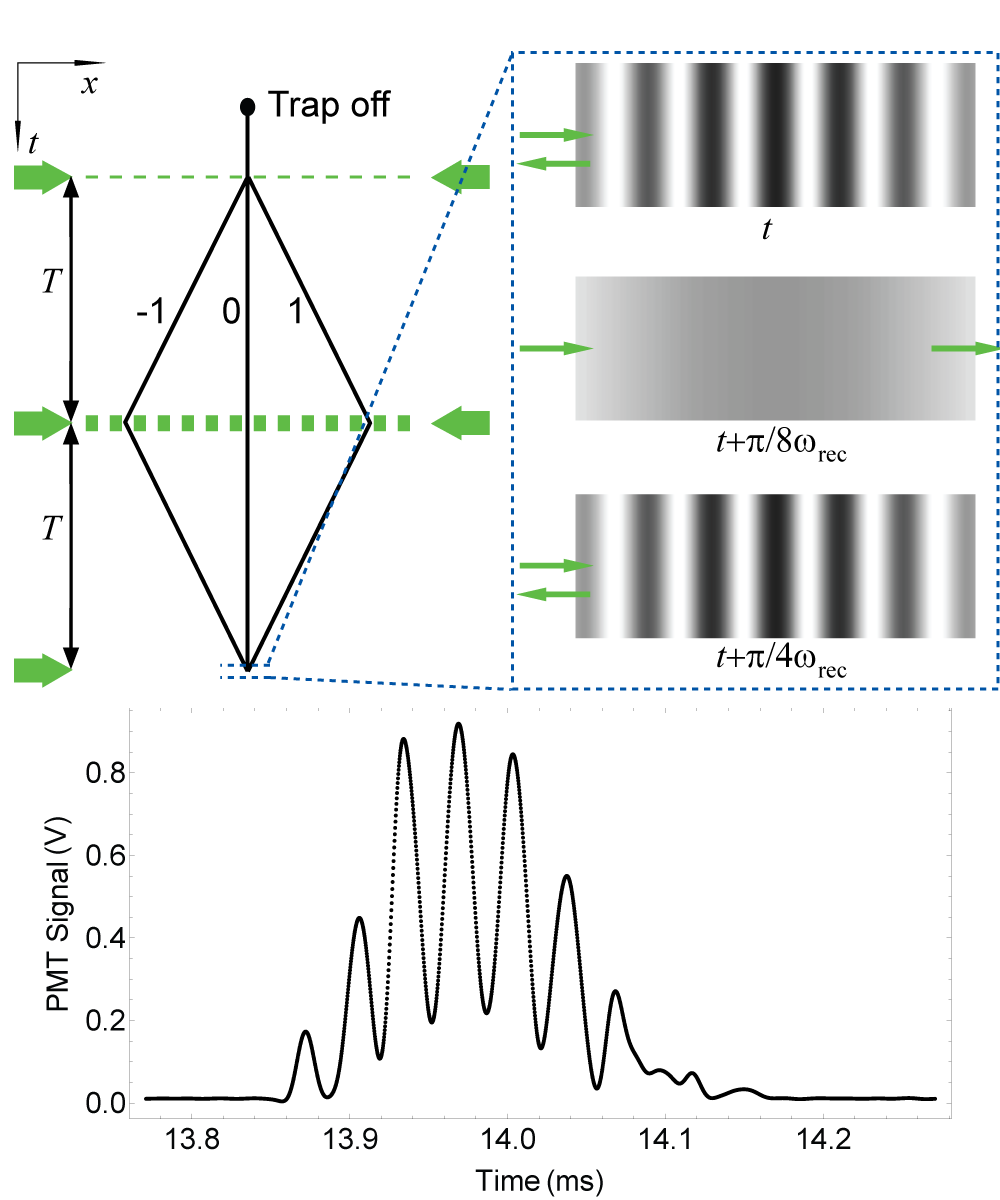}
\caption{(color online) Contrast interferometer geometry and readout. On the top left, the geometry for the contrast interferometer is depicted. The splitting pulse diffracts the BEC into three branches, at $t=0$. A mirror pulse at $t=T$ reverses the momenta of the moving branches. To the right, the atomic density patterns seen at three times near $t=2T$ demonstrate the source of the oscillating back-reflection signal. Light is Bragg reflected from a high-contrast atomic density grating, while it passes through a uniform density cloud a short time later. At the bottom, a sample readout signal from a single run of the interferometer, with $T = 7\;{\rm ms}$, is shown.}
\label{fig:Fig1}
\end{figure}
Specifically, we present a contrast interferometer (CI) using a $^{174}\rm Yb$ BEC as source. The CI is a promising design for precision measurements of the fine structure constant\cite{gupta02}. We demonstrate interferometer times as long as $22\; {\rm ms}$, more than three times longer than previous CIs. We show that the interferometer signal quality does not degrade over such long times, even without vibrational isolation---a dramatic improvement over traditional interferometers\cite{hensley01}. Diffraction phases due to pulses far from the Raman-Nath (short pulse) regime are an important systematic effect for high-precision interferometers utilizing large-momentum transfer beam splitters\cite{muller08b,chiow11}. We report the first measurement of diffraction phases for such pulses, successfully modeling and correcting for this effect. Finally, we demonstrate quantitative control over atomic interaction effects within our BEC source. Controlling these interactions is important for achieving high accuracy measurements with BEC interferometers. 

The paper is organized as follows. In section \ref{sec:Advances} we discuss technical advances in the interferometer itself: the Yb source and increased interferometer time. In section \ref{sec:DiffPhase}, we discuss the diffraction phases model and its successful implementation into our data analysis. In section \ref{sec:AtomInt}, we discuss the physics of atomic interactions and test our models against experiments. Finally, in section \ref{sec:future}, we consider scaling of precision and a variety of systematic effects to assess the possibility of a future sub-ppb measurement of the Yb photon recoil frequency and the fine structure constant.

\section{Interferometer Advances}
\label{sec:Advances}
\subsection{Yb Contrast Interferometer}
In a CI, a cold cloud is released from a trap and allowed to expand for some time. Then, at $t=0$, the atomic wave functions are diffracted into three branches (see Fig. \ref{fig:Fig1}) by a short standing-wave pulse of light, the ``splitting pulse.'' A simple model treats the three branches as plane-wave states with equal densities and momenta $-2\hbar k$, $0$, and $2\hbar k$, where $k=2\pi/\lambda$ is the wavenumber of the light used to make the diffraction gratings.

At $t=T$, a longer pulse, the ``mirror pulse,'' is used to reverse the momenta of the two moving branches. Finally, at $t=2T$, a traveling light pulse probes the cloud. Interference between the three momentum states creates a grating of atomic density with contrast that rises and falls over time. The grating period, $\lambda/2$, causes the traveling pulse to coherently back-reflect when the grating has high contrast and pass through when it has low contrast\footnote{Interference between the $\pm 1$ branches also creates a grating of spacing $\lambda/4 $, which our readout technique is insensitive to.}. The contrast of the density grating, and thus the reflected light signal intensity, oscillates as 
\begin{equation}
\label{CIphaseCombo}
A(t)\cos^2\left( \frac{\phi_1(t) + \phi_{-1}(t)} {2} - \phi_0(t) \right),
\end{equation}
where $A(t)$ is an amplitude envelope caused by the finite spatial extent of the initial condensate, as opposed to the infinite plane waves of the simple model, and $\phi_i(t)$ is the time-dependent phase of the $i$ branch. This signal oscillates due to the kinetic energy of the $\pm 1$ branches. An example signal is shown at the bottom of Fig. \ref{fig:Fig1}. We measure the phase of the signal at time $2T$, $\phi(2T)=8\omega_{\rm rec} T+\phi_{\rm offset}$, where $\omega_{\rm rec} = \hbar k^2/(2m)$ is the recoil frequency, $m$ is the mass of a single atom, and $\phi_{\rm offset}$ contains a number of phase shifts common to interferometers of differing $T$. Measuring $\phi(2T)$ for two different values of $T$ allows high-precision measurement of $\omega_{\rm rec}$. This may then be combined with other fundamental constants to arrive at the fine structure constant\cite{weiss93}.

The experiments reported in this paper were performed in the apparatus described in \cite{hansen12}. All experiments used the isotope $^{174}{\rm Yb}$, which has a recoil frequency $\omega_{\rm rec} = 2\pi\times 3.7\;{\rm kHz}$. For this work, we produced $^{174}\rm Yb$ BECs of approximately 150,000 atoms with a cycle time of $15 \; {\rm s}$. We also verified that a CI signal could be obtained with a non-degenerate (i.e., above the condensation temperature) source. However, the signal quality was substantially inferior to that obtained from a BEC source.

We used a single, retro-reflected laser beam to create the standing-wave gratings. The beam and the retroreflection both had Gaussian profiles with $3\;{\rm mm}$ waists. For the readout light, we used a separate beam of much smaller waist ($\approx 200\;{\rm \mu m}$). All of the beams were oriented horizontally. The laser frequency was detuned from the $556\;{\rm nm}$ intercombination line ($^1S_0\rightarrow {^3P_1}$) by $\approx 450 \Gamma$, where $\Gamma = 2 \pi \times 182 \;{\rm kHz}$ is the natural linewidth. Some data sets were taken with blue detuning and others with red detuning. No substantial differences were found between the two as, even at our peak $t=0$ densities of $9\times 10^{13}\; {\rm cm}^{-3}$, index of refraction effects fall below our level of sensitivity. For the readout light, the detuning was reduced to $50 \Gamma$. A substantial technical advantage of using such a narrow transition is that all of these frequencies can be accessed with a single $200\;{\rm MHz}$ acousto-optic modulator. 

For the splitting pulse we used a square pulse of length $1.5\;{\rm \mu s}$. The mirror pulse was a Gaussian with intensity $1/e$ half-width $30\;{\rm \mu s}$. This pulse achieved a second-order diffraction efficiency of $90\%$. The readout pulse was typically a $180\;{\rm \mu s}$ square pulse. 

\subsection{Large $T$ and Vibration Insensitivity}
\begin{figure}
\centering
\includegraphics[width=0.9\linewidth]{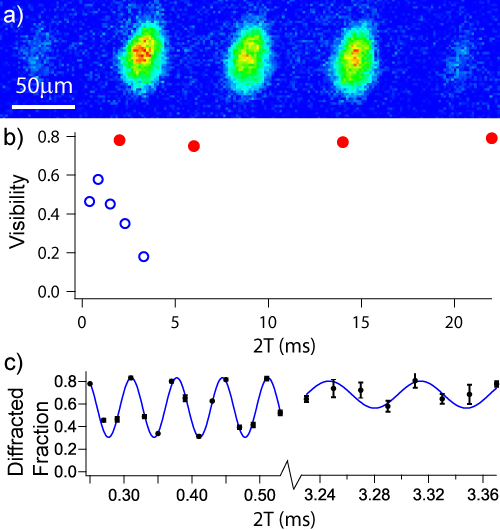}
\caption{(color online) Vibration insensitivity. In a) an absorption image made at the time of the mirror pulse demonstrates the clear separation of the arms in a $T=11\;{\rm ms}$ CI. In b) the blue, open circles indicate the visibility of a traditional interferometer, which drops off strongly around $T=1.5\;{\rm ms}$, due to vibrations. The red, filled circles show the CI visibility measured in the same apparatus. (Error bars are smaller than the markers.) For all CI data sets, we see sample standard deviations in the phase between 130 and 190 mrad, with no clear trend versus $T$. Part c) shows signals from the traditional interferometer as fraction of atoms in the $\pm 2\hbar k$ states after the final pulse. These show the fall-off in visibility with increasing $T$.}
\label{fig:Fig3}
\end{figure}

Most interferometer geometries use an externally applied diffraction grating---either material gratings or standing waves of light---as a ruler to read out the final phase. In a CI, two matter-wave gratings are produced which serve as phase rulers for one another. This eliminates the requirements for complex control of the external grating. For a light-pulse interferometer this involves, at minimum, active vibration control of optics\cite{hensley01}. This insensitivity of the CI to vibrations has allowed us to extend the interferometer time to $22\;{\rm ms}$ without any vibration control. Figure \ref{fig:Fig3}a shows an absorption image of the BEC at the time when the mirror pulse is applied for such an experiment, demonstrating the spatial separation $( 90\;{\rm \mu m})$ between the nearest neighboring coherent branches of the wave function. We see no loss of visibility or signal-to-noise ratio over this time (see Fig. \ref{fig:Fig3}b). The interferometer time in this work was limited purely by the atoms falling out of the diffraction beams. In the future, this can be mitigated with larger beams or moving to a vertical interferometer geometry. 

To compare the CI to a more traditional atom interferometer, we created an identical, three-branch interferometer, but with an external diffraction grating readout. The weak traveling-wave pulse at the end of the CI was replaced by a short standing-wave pulse identical to the splitting pulse. The populations in the $\pm 2 \hbar k$ states oscillate at $4 \omega_{\rm rec}$ as the time of the final pulse is scanned. We observe this oscillation in time-of-flight absorption images. As seen in Fig. \ref{fig:Fig3}b, the visibility of this signal begins to decline sharply around $T=1.5\;{\rm ms}$, while the CI signal continues essentially unchanged to $T = 11\;{\rm ms}$.

The current state of the art for recoil measurements is part per billion (ppb) accuracy\cite{bouchendira11}. Controlling unwanted interactions is key to this level of accuracy. The symmetry of the CI geometry controls several external perturbations. The phase measured in a CI, eq (\ref{CIphaseCombo}), is insensitive to any external field that causes a constant shift of energy or an energy gradient across the interferometer. Given the small volume sampled by the CI, and Yb's lack of magnetic moment, the effects of external fields can be reduced or measured and subtracted to below the ppb level, as will be described in Section \ref{sec:future}.

We now turn, instead, to the two largest systematic shifts for a fine structure constant measurement with a Yb CI: diffraction phases and atomic interactions. The immunity of a CI to outside influence allows us to cleanly probe and control these effects.

\section{Diffraction Phases}
\label{sec:DiffPhase}
Diffraction phases were first suggested as an important systematic effect in a CI by Buchner et al. \cite{buchner03}. They have been studied previously with diffraction from material gratings\cite{perreault06} and from light-pulse gratings near the Raman-Nath regime ($\tau \ll 1/\sqrt{\Omega_{\rm R}\omega_{\rm rec}}$, where $\tau$ is the length of the pulse and $\Omega_{\rm R}$ is the Rabi frequency for one of the diffraction beams) \cite{yue13}. In this regime, the diffraction phases are essentially unaffected by small fluctuations in intensity or pulse duration between shots and so cancel in the final analysis.

While our splitting pulse falls in the Raman-Nath regime, the mirror pulse falls between it and the easy to calculate Bragg ($\tau \gg \pi/2\omega_{\rm rec}$) regime\cite{gupta01}. Pulses in this intermediate regime are critical elements to recoil measurements both as mirror pulses and for acceleration of moving interferometer branches. In this regime, diffraction phases depend sensitively on small changes of intensity or pulse shape from shot to shot. To correct for these changes, we record the time-dependent intensity of each mirror pulse.

We treat the laser pulses as single-atom effects, ignoring collective effects and energy shifts due to inter-atomic interactions. For our detunings and the reduced atomic density at the time of the mirror pulse, collective effects (e.g., super-radiance\cite{inouye99}) are negligible. The single particle Hamiltonian is
\begin{equation*}
\frac{p^2}{2m}+\hbar \omega_0 \ket{e}\bra{e}\\
+e^{-i\omega t}\ket{e}\bra{g}\left(\Omega_1 e^{ikx}+\Omega_2 e^{-ikx}\right)+{\rm h.c.} ,
\end{equation*}
where $\omega_0$ is the energy difference between ground ($\ket{g}$) and excited ($\ket{e}$) electronic states, $\omega$ is the angular frequency of the light, and $\Omega_i$ is the Rabi frequency of the $i^{\rm th}$ diffraction beam. We calculate the effect of each pulse numerically, using the measured time dependence of the intensity\footnote{This procedure will be described in a forthcoming publication.}. For simulation purposes, the intensity at the position of the BEC is needed, whereas only the total power in the diffraction beams can feasibly be recorded. Calibration of local intensity for a given power is obtained by recording the effect of mirror pulses with varying powers.

\begin{figure}
\centering
\includegraphics[width=0.9\linewidth]{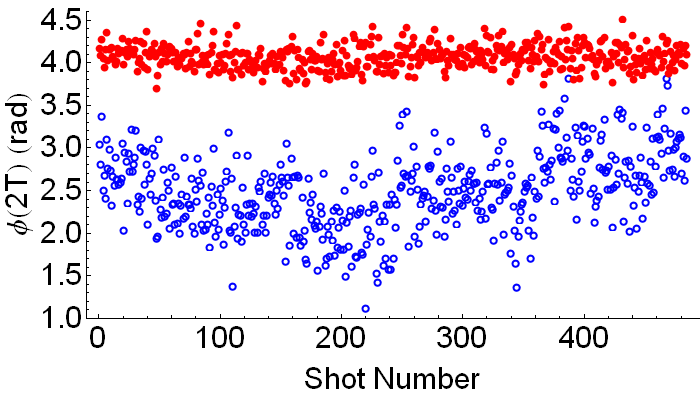}
\caption{(color online) Diffraction phase correction. The unfilled, blue circles show the phase of the CI signal before any systematic corrections. The filled, red circles show the same data after correcting for the diffraction phase with a numerical model. Both short-time jitter and hours-scale intensity drifts are corrected by the diffraction phase analysis.}
\label{fig:Fig4}
\end{figure}
After calculating the diffraction phase we subtract it from the phase of the contrast signal, $\phi(2T)$. We extract $\phi(2T)$ by fitting a sine wave to the signal and finding its phase at time $2T$. Figure \ref{fig:Fig4} shows data before and after this correction. In addition to a uniform noise width, we also see a marked drift in phase over time. This drift is eliminated in the corrected data, showing that it arose from drifts in laser intensity. We also note that the noise width is reduced by the correction. Overall, the standard deviation of the data set drops from $440\;{\rm mrad}$ to $140\;{\rm mrad}$. The ability to correct for phase shifts induced by both laser drift and random laser noise gives good confirmation of our model. All data discussed below have these corrections applied.

Experiments with different $T$ have the atoms in different parts of the diffraction beam when the mirror pulse occurs. Thus, local intensity will differ across $T$, even for pulses with identical power and temporal shape. Substantial differences in diffraction phase between experiments at different $T$ can accrue from these local intensity differences. So, the diffraction phase is an important systematic effect even with perfect control of laser pulse powers. Our simulations can correct for this shift to the $0.1\;{\rm mrad}$ per pulse level, limited by the accuracy of the power to local intensity calibration.

\section{Atomic Interactions}
\label{sec:AtomInt}
Finally, we consider atomic interactions. We use the mean-field approximation, wherein all atoms are assumed to have the same single-particle wave function. Each atom feels an effective potential due to the other atoms in the BEC. The potential can be parametrized purely by the $s$-wave scattering length $a_s$. This approach leads to the Gross-Pitaevskii equation (GPE):
\begin{equation}
i\hbar \frac{\partial \psi}{\partial t} = -\frac{\hbar^2}{2m} \nabla^2 \psi +  \frac{4\pi \hbar^2 N_{\rm at} a_s}{m}|\psi|^2\psi,
\end{equation}
where $\psi$ is the mean-field wave function. In this work, our Yb BECs typically contain $N_{\rm at} \approx 150,000$ atoms. This large number coupled with the small scattering length of $^{174}{\rm Yb}$ ($a_s = 5.6\;{\rm nm}$) make the GPE accurate to better than 1\%. 

\begin{figure}
\centering
\includegraphics[width=0.9\linewidth]{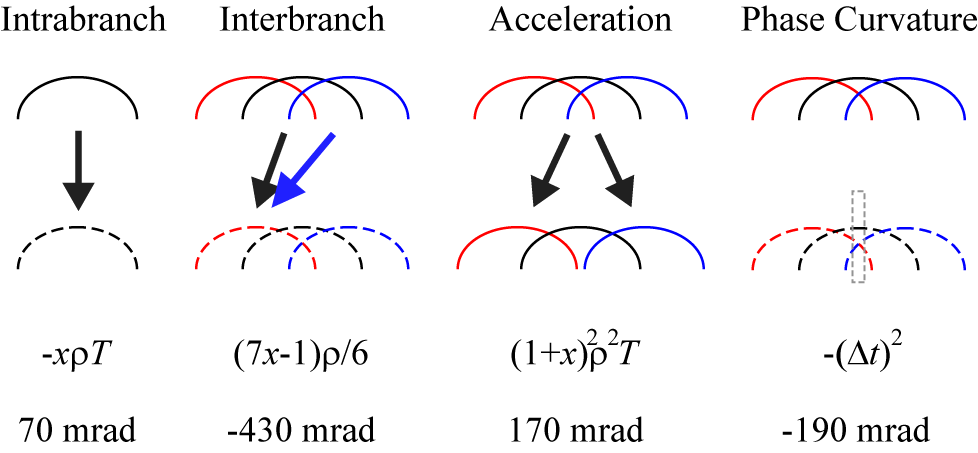}
\caption{(color online) Interaction effects. The four important interaction effects are illustrated. Below each illustration are the scaling of the effect with splitting parameter, $x$, density at time of splitting, $\rho$, the difference between $2T$ and the time of perfect overlap, $\Delta t$, and $T$. The branches are represented by red, black, and blue lumps. Solid lines show density profiles, and dashed lines show phase profiles. The gray box in the phase curvature illustration highlights that the center of the black $0\hbar k$ branch interferes with the wings of the $\pm 2 \hbar k$ branches. Values given to illustrate the relative sizes are for an experiment with $T=11\;{\rm ms}$, 2 ms of expansion time, and $x=0.01\pm 0.01$.}
\label{fig:fig5}
\end{figure}

To simplify the GPE, and to clarify the physics, we combine scaling solutions and the slowly-varying envelope approximation (SVEA)\cite{jamison11}. These techniques allow each of the branches populated by the splitting pulse to be treated independently. Each momentum branch of the condensate then obeys a GPE with extra interaction terms to describe interactions between branches. The phase evolution depends on the ratio in which atomic density splits between the various branches. We introduce an asymmetry parameter, $x$, such that the ratio of densities is $1-x:1+2x:1-x$ for the $-2\hbar k, 0,$ and $2\hbar k$ branches. The four interaction effects seen in the SVEA are summarized in Fig. \ref{fig:fig5}. 

First, there are intrabranch energy shifts. These shifts arise from the interaction energy of a single branch of an atom with the total atomic density in that branch of the interferometer. 
From equation \ref{CIphaseCombo}, the phase of the signal will be shifted by the difference between intrabranch energy for the moving versus non-moving branches. Thus, the phase shift is proportional to $-x \rho T$, where $\rho$ is the atomic density just before the initial splitting. 

There are similar interbranch interactions. During the time the branches of the interferometer are overlapped in space the $2 \hbar k$ branch of an atom will interact with the total atomic density in the $0$ branch of the BEC. This gives an energy shift analogous to that from intrabranch interactions. Similar shifts arise for all other pairs of momentum states. Importantly, interbranch interactions are twice as strong due to the distinguishability of the two branches. The moving branches have a shift proportional to $\rho ((1+2x)+0.5(1-x))/3$ (the $0.5$ comes from the moving branches overlapping with one another half the time they overlap with the nonmoving branch) while the shift on the non-moving branch is proportional to $\rho(2)(1-x)/3$. Thus, the overall effect scales as $\rho(7x-1)/6$.

Additionally, as the branches separate they exert forces on each other that accelerate the moving branches. This increases the momentum from $t=0$ to $t=T$ by an amount $\hbar \Delta k \propto (1+x) \rho$. However, the mirror pulse is not exactly a mirror. Rather, it changes the momentum by $\pm 4 \hbar k$. The branch with momentum $\hbar(2k + \Delta k)$ just before $t=T$ will have momentum $\hbar (-2k + \Delta k)$ just after $t=T$. Thus, the term $4\hbar^2 k \Delta k$ will cancel out of the total phase accumulation. Only the term quadratic in $\Delta k$ from the total kinetic energy survives. Thus, the total phase shift from this effect is proportional to $(\Delta k)^2 T \propto (1+x)^2 \rho^2 T$.  

A final, less obvious effect of interactions involves the phase curvature across the condensate. Due to the acceleration effect, the branches may not be perfectly overlapped at time $2T$. The actual time of perfect overlap is referred to as the closing time of the interferometer. In this case, the grating is formed by the interference between non-analogous parts of the different branches. In figure \ref{fig:fig5} the gray box highlights the fact that the center of the $0$ momentum branch interferes with the wings of the $\pm 2 \hbar k$ branches. The phase accumulated due to interactions in the BEC before splitting is curved like the density profile that generates it, as first demonstrated in \cite{hagley99}. Therefore, the phase is greatest in the center of a branch and decreases into the wings. This effect scales like $(\Delta t)^2$, where $\Delta t$ is the difference in time between the proper closing time and the time when data is taken. This shift can be made negligibly small by taking data at $\Delta t = 0$ rather than at $2T$. Taking data at $\Delta t = 0$ spoils the cancellation of the $4\hbar^2 k \Delta k$ term in the acceleration shift. Thus, a trade-off must be made in deciding between $2T$ and the closing time. In this work all data was taken around $2T$.

To test the accuracy of these calculated corrections, we apply them to two $T = 11\;{\rm ms}$ data sets differing only in their density splittings, one with $x = 0.29 \pm 0.01$ and another with $x = -0.14 \pm 0.01$. In these experiments, the BEC is allowed to expand for only $2\; {\rm ms}$ before the splitting pulse. For a high-precision data set, the expansion time may be $10\;{\rm ms}$, which reduces the density, and thus the interaction effects, by a factor of more than 10. The artificially short expansion time, and resultant high density, used for these two data sets magnifies the interaction effects. Before applying interaction shift corrections, their phase difference is $0.70 \pm 0.03\;{\rm rad}$. After applying the corrections, the difference is $0.02 \pm 0.1\;{\rm rad}$. The large error bar is due to uncertainty in trap geometry and turn-off. In a high-precision recoil experiment, trap parameters can be both better controlled and better measured, reducing the uncertainty. Together with longer expansion times, these should enable correction of interaction effects at the $<5\;{\rm mrad}$ level.

\section{Future Prospects}
\label{sec:future}
We apply the diffraction phase and atomic interaction corrections described in Sections III and IV to all our data. On separate data runs, we have achieved accuracies of 45 and 60 ppm in the Yb recoil frequency (which translates into 23 and 30 ppm in alpha, respectively). Unfortunately, uncertainty in trap shape and turnoff limits our current accuracy. Comparing larger data sets taken on different days allows us to achieve higher precision. However, these comparisons show greater inaccuracy indicating that there are experimental parameters that drift from day to day. Better measurement and control of all experimental parameters such as trap shape will be key to future work, but should not constitute a substantial impediment.

To assess the scalability of our interferometer, we have made two separate measurements of the Yb recoil frequency with 7 ppm precision, one with $\Delta T = 6\; {\rm ms}$ (uncertainty in $\phi(2T)$, $\delta \phi = 7\;{\rm mrad}$ in 500 runs) and a second with $\Delta T = 10\; {\rm ms}$ ($\delta \phi = 12\;{\rm mrad}$ in 150 runs) by combining data sets from different days. We use these precision benchmarks to discuss scaling to higher precision through acceleration of the moving branches.

Competitive recoil measurements will require acceleration of the moving branches of the CI to momenta $\pm 2N\hbar k$, where $N$ is an integer. As the phase evolution in the CI scales with $N^2$\cite{gupta02}, achieving $N=100$ would allow a precision of 0.7 ppb simply by scaling current results. At this level, improved tests of QED could be made\cite{aoyama12,hanneke08,rana12}. Achieving $N=100$ requires reasonable extension of previously demonstrated techniques for large coherent accelerations, using either Bloch oscillations\cite{muller08b,denschlag02} or a sequence of Bragg pulses\cite{chiow11}. The diffraction phase systematic scales like $N^{1/2}$\footnote{This scaling assumes each diffraction pulse is independently calibrated. See [23]}. The interaction systematic does not scale with $N$ at all. Therefore, our demonstrated control of these effects is encouraging for the prospect of achieving sub-ppb accuracy in an $N=100$ experiment.

Systematic effects can be organized into three groups: Interactions of atoms with external fields, the diffraction laser beams, and each other. Having addressed atom-atom interactions in Section \ref{sec:AtomInt}, we consider the sizes of external field interactions and laser beam interactions other than the diffraction phase, which was dealt with in Section \ref{sec:DiffPhase}. The results are summarized in Table \ref{tab:SystematicShifts}.

The phase of the signal depends on the combination of phases seen in \eqref{CIphaseCombo}. Thus, only interactions which cause a curvature of phase evolution across the interferometer can actually shift the measured phase. As mentioned above, a precision of 0.7 ppb in $\omega_{\rm rec}$ can be reached with our current phase precision of $12\;{\rm mrad}$ in a $T=10\;{\rm ms}$ CI, if acceleration up to $N=100$ is added. As a benchmark for accuracy, we will discuss systematic shifts at the $1\;{\rm mrad}$ level for such a CI, which corresponds to $<0.1\;{\rm ppb}$ in $\omega_{\rm rec}$.

External magnetic fields have no effect on the ground state of $^{174}{\rm Yb}$. They can shift the excited state energy and thus potentially affect the laser beam interactions. However, the linearly polarized light used for the diffraction pulses can be decomposed into equal magnitudes of right and left circularly polarized light. This shows that an applied magnetic field which splits the excited $^3 P_1$ state will, to leading order, have no net effect, as one of $m =\pm 1$ will shift away from resonance while the other shifts toward. To reduce the residual shift of the diffraction phase (largest laser beam interaction shift) to the $1\;{\rm mrad}$ level, the difference in magnetic field between runs of different $T$ must be $\le 1\;{\rm G}$. Magnetic field shifts larger than this would substantially harm BEC production, likely leading to a complete loss of atoms. Thus, such large magnetic field shifts would be readily detectable\footnote{In fact, ambient magnetic field fluctuations and long-term drifts have been measured in the lab space by a separate ion trap experiment\cite{hoffman13}, showing field excursions to be $<10\;{\rm mG}$}.

External (quasi-static) electric fields can affect the interferometer through the Stark shift. The energy of an atom in a static electric field is $E_{\rm Stark} = p{\cal E}^2$, where $p$ is the atomic polarizability\cite{beloy12,safronova12} and $\cal E$ is the magnitude of the electric field. The curvature of the energy along the interferometer axis, $z$, gives 
\begin{equation*}
\Delta E_{\rm Stark} = p\left[\left(\frac{\partial {\cal E}}{\partial z} \right)^2+{\cal E}\frac{\partial^2 {\cal E}}{\partial z^2}\right]\left(\Delta z \right)^2.
\end{equation*}
Treating the closest possible charge (accumulated charge on the vacuum viewports, $5\;{\rm cm}$ from the atoms) as a point source, we find the charge must be no larger than $0.3\;{\rm nC}$ to keep the differential Stark shift to $1\;{\rm mrad}$. A charge this large could easily be detected with an electrometer\footnote{For example, the Standard Imaging SuperMAX Electrometer can measure charges as small as $1\;{\rm fC}$}. The Stark shift due to blackbody radiation can be considered as a quasi-static effect as well. The variation of the black body field in a vacuum chamber is far too small to have a noticeable impact on the CI phase\cite{safronova12}.

The CI is sensitive to curvature of the gravitational potential, which corresponds to gradients in the acceleration due to gravity, $g$. To subtract the shifts due to gravity, the gradient of $g$ must be measured to an accuracy of $3\times 10^{-6}\;{\rm s^{-2}}$. Commercial gradiometers can measure at least two orders of magnitude better than this\footnote{For example, the Gravitec gravity gradiometer has a sensitivity of $5\times 10^{-9}\;{\rm s^{-2}}$ for measuring static gradients. }.

The remaining potential shifts are due to the geometry of the diffraction beams and to index of refraction shifts of the recoil momentum. In a future experiment, we plan to increase the beam waists for the diffraction beams to $w = 8\;{\rm mm}$. For beams this size, the wavefront curvature and Guoy phase combine to give a shift of the momentum per photon of $-4\;{\rm mrad}$ \cite{clade06}. As this shift is always negative and is a well-known function of beam waist, some of this shift can be corrected for. Finally, the long Rayleigh range $z_{\rm R} = \pi w^2/\lambda = 360\;{\rm m}$ makes the relative intensity variation $5\times 10^{-10}$ over the $8\;{\rm mm}$ peak path separation. For our planned beam intensity and an illumination time of $1\;{\rm ms}$ at this separation, we find a shift of $\approx 10 \;{\rm nrad}$.

Deviation of the two diffraction beams from perfect counter propagation reduces the net momentum transfer to $2k\cos(\delta)$ where $\delta$ is the half-angle between the beams. By carefully coupling the beams into one another's single mode fibers, the deviation can be constrained to roughly $\delta = 3\times10^{-5}$\cite{clade06}. This leads to a shift of $-7\;{\rm mrad}$. Like the wavefront curvature and Guoy phase, this shift is always negative and has an easily fit functional form. Therefore, part of this shift can be measured and corrected.

The index of refraction, $n$, only affects the initial splitting pulse, as the other pulses cause $\approx 100\%$ population transfer for each of the three spatially separated branches of the interferometer\cite{campbell05}. By the same argument given above for the interaction induced momentum kick, the index of refraction shift is only proportional to the square of the momentum shift.

In our current experiments with an expansion time of $10\;{\rm ms}$ before the splitting pulse we have $n-1 = -3\times 10^{-5}$. This shifts the total phase by $4\omega_{\rm rec}(n-1)^2(2T)$. With $N=100$, this is a fractional shift of magnitude $9\times 10^{-14}$, or $2\;{\rm \mu rad}$. 

\begin{table}
\begin{tabular}{|p{4.5cm}|c|}
\hline \centering \textbf{ Interaction} & \textbf{Relative Shift} (ppb) \\ 
\hline  Electric Fields & $<0.1$ \\ 
\hline  Magnetic Fields &  $<0.1$ \\ 
\hline  Blackbody Radiation &  $<0.1$ \\ 
\hline  Gravity Gradients &  $<0.1^*$ \\ 
\hline  Wavefront Curvature and Guoy Phase &  $<0.2^*$ \\ 
\hline  Beam Alignment &  $<0.4^*$ \\ 
\hline  Index of Refraction &  $<0.1$ \\ 
\hline 
\end{tabular} 
\caption{Systematic shifts of $\omega_{\rm rec}$. The relative size of systematic shifts discussed in the text are summarized. Values with an $^*$ reflect size after corrections described in the text. For the experiment discussed, $1\;{\rm ppb}$ would correspond to a $18\;{\rm mrad}$ shift.}
\label{tab:SystematicShifts}
\end{table}

In conclusion, we have demonstrated the first matter-wave interferometer with Yb. We have substantially extended the reach of contrast interferometry by demonstrating experimental times more than three times longer than previously possible and explicitly showing immunity to vibrations. Combining the CI geometry and the Yb atom controls or eliminates a number of potential systematic shifts for a future sub-ppb-level recoil measurement. Finally, we have shown control over the two largest remaining systematic effects for recoil measurements with a CI. These effects are relevant to other high-precision atom interferometers, as well: Diffraction phases are a concern for any interferometer with large-momentum beam-splitting via multiple transitions without changing internal state, and atomic interactions are a concern for any BEC interferometer. The level of control demonstrated  in this work is encouraging for the prospect of achieving sub-ppb precision and accuracy with a Yb BEC CI.

This work was supported by the NSF and by a NIST Precision Measurement Grant. We thank the other members of the ultracold atoms group at UW for technical assistance.

\end{document}